\documentclass[aps,twocolumn,secnumarabic]{revtex4-1}
\usepackage{amsmath}
\usepackage{graphicx}      
\usepackage{bm}            

\begin{document}

\title{Simplified Chirp Characterization in Single-shot \\Supercontinuum Spectral Interferometry}
\author {DinhDuy Tran Vu$^1$}
\author {Dogeun Jang$^1$}
\author {Ki-Yong Kim$^1$}
\altaffiliation{kykim@umd.edu}
\affiliation{$^1$ Institute for Research in Electronics and Applied Physics, University of Maryland, College Park, MD, 20742}

\begin{abstract}
	\noindent
	Single-shot supercontinuum spectral interferometry (SSSI) is an optical technique that can measure ultrafast transients in the complex index of refraction. This method uses chirped supercontinuum reference/probe pulses that need to be pre-characterized prior to use. Conventionally, the spectral phase (or chirp) of those pulses can be determined from a series of phase or spectral measurements taken at various time delays with respect to a pump-induced modulation. Here we propose a novel method to simplify this process and characterize reference/probe pulses up to the third order dispersion from a minimum of 2 snapshots taken at different pump-probe delays. Alternatively, without any pre-characterization, our method can retrieve both unperturbed and perturbed reference/probe phases, including the pump-induced modulation, from 2 time-delayed snapshots. From numerical simulations, we show that our retrieval algorithm is robust and can achieve high accuracy even with 2 snapshots. Without any apparatus modification, our method can be easily applied to any experiment that uses SSSI.
\end{abstract}

\maketitle

\section{Introduction}
\noindent
Single-shot spectral interferometry (SSI) is an ultrafast optical method that can measure ultra-rapid refractive index transients induced by ultrashort laser pulses \cite{re1,re2}. This measurement can provide a direct view of how a laser-induced perturbation evolves in time and space in a single-shot. In this technique, a pump pulse induces a refractive index transient in a medium, and a chirped reference and a time-delayed replica (probe) pulse, upon which the pump-pulse-induced phase shift has been imposed, interfere in an imaging spectrometer, producing a spectral interferogram. The corresponding spatiotemporal (time and 1-dimensional space) evolution of refractive index transient is then reconstructed with frequency-to-time mapping or full Fourier transform methods \cite{re2}. In particular, supercontinuum (SC) light has been used for the reference and probe pulses to provide a temporal resolution better than $\sim$10~fs \cite{re2}. This single-shot supercontinuum spectral interferometry (SSSI) and SSI techniques have been successfully applied to capture laser-induced double step ionization of helium \cite{re3}, laser-heated cluster explosion dynamics \cite{re4}, laser wakefields \cite{re5}, optical nonlinearity near the ionization threshold \cite{re6}, and electronic and inertial nonlinear responses in molecular gases \cite{re7,re8,re9,re10}.  SSI has been also used to capture terahertz waveforms in snapshots without pump-probe scanning \cite{re11,re12}.

Unlike self-referencing nonlinear diagnostics such as FROG \cite{re13,re14} and SPIDER \cite{re15,re16}, linear spectral interferometery including SSSI requires pre-characterized reference/probe pulses prior to its use. One method to pre-characterize a chirped probe in SSSI is to scan the delay between the pump-probe pulses while tracking a characteristic central extremum in the modulated probe phase or spectrum \cite{re2}. This method can determine the spectral phase of SC probe light to arbitrary order \cite{re17}. However it relies on stationary-phase and small-perturbation approximations, which can be problematic when the phase modulation is too large or too asymmetric. In addition, this requires repetitive measurements over the entire chirp window for an accurate extraction of higher order dispersion coefficients. In particular, for a SC probe that has an extremely large bandwidth, a large number of pump-probe scans are necessary. This scanning method is impractical for use with low-repetition-rate laser sources.

For these reasons, we aim to develop a method that can simplify or avoid the pre-characterization process if possible and potentially to characterize both SC reference/probe pulses and pump-induced transients with a minimal number of repetitions.

\section{Background theory}
\begin{figure}[htbp]
	\centering
	\includegraphics[width=0.9\linewidth]{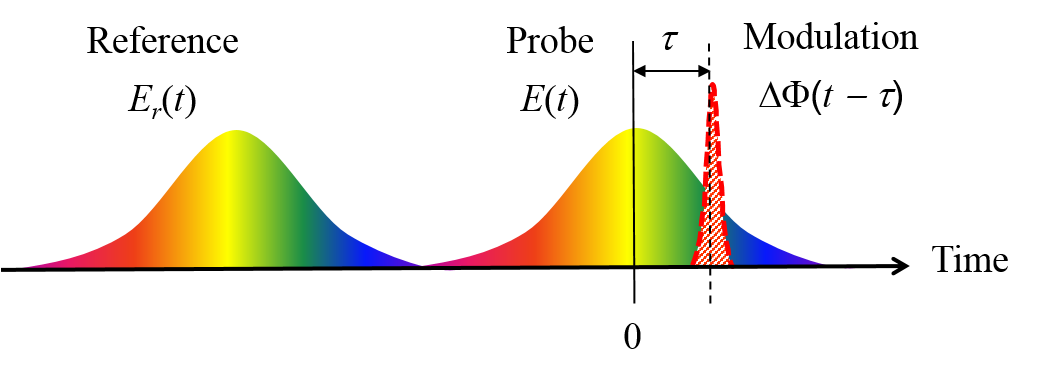}
	\caption{Schematic of SSSI consisting of a chirped supercontinuum (SC) pulse (reference, $E_r (t)$) and a time-delayed replica pulse (probe, $E (t)$) upon which a pump-induced ultrafast modulation $\Delta\Phi(t-\tau)$ has been imparted at a time delay $\tau$ with respect to the probe.}
	\label{Fig1}
\end{figure}

\noindent
In SSSI, the probe SC pulse, $E (t)$, is perturbed by a pump-induced modulation, $\Delta\Phi(t-\tau)$, applied at a time delay $\tau$ with respect to the probe pulse (see Fig.~\ref{Fig1}). The reference SC pulse, $E_r (t)$, precedes the pump in time and thus remains unaffected.

The perturbed probe pulse $\overline{E}(t)$ can be written as
\begin{equation}
	\overline{E} (t)=E (t) e^{i\Delta\Phi(t-\tau)},
\end{equation}
where $E (t)$ is the unperturbed probe pulse. Then  $\Delta\Phi(t)$, the same pump-induced modulation but applied at $\tau = 0$, can be extracted from the interference between the reference and probe pulses in the frequency domain as
\begin{equation}\label{Eq2}
	\Delta \Phi (t)=-i\ln \left[ \frac{{{F}}\left\{ \left| {\overline{E}}(\omega ) \right|{{e}^{i(\Delta {{\varphi }_{\tau }}(\omega )+\varphi (\omega ))}}{{e}^{-i\omega \tau }} \right\}}{{{F}}\left\{ \left| {{E}}(\omega ) \right|{{e}^{i\varphi (\omega )}}{{e}^{-i\omega \tau }} \right\}} \right],
\end{equation}
where $F$ denotes the Fourier transform, $|\overline{E}(\omega)|$ and $|E(\omega)|$ is the spectral amplitude of the perturbed and unperturbed probe pulses, respectively, that can be directly measured by a spectrometer; $\Delta\varphi_\tau (\omega)$ is the phase difference between the modulated probe and reference pulses at a given $\tau$ that can be obtained from an interferometer; and $\varphi(\omega)$ is the spectral phase of the unperturbed probe (or reference) pulse. Here to retrieve $\Delta\Phi(t)$, the spectral phase $\varphi(\omega)$ needs to be characterized. In general, the spectral phase of a chirped pulse can be expressed in a Taylor expansion around the central frequency $\omega_c$ as
\begin{equation}
\begin{split}
  	\varphi (\omega )&=\varphi_0+b_1(\omega-\omega_c) \\
  	  & \quad +{{b}_{2}}{{(\omega -{{\omega }_{c}})}^{2}}+{{b}_{3}}{{(\omega -{{\omega }_{c}})}^{3}}+...,
\end{split}
\end{equation}
where $\varphi_0$ is the absolute spectral phase; $b_1$ is the first order dispersion coefficient related to a pulse shift in time; $b_2$ and $b_3$ are the second and third order dispersion coefficients, respectively. Here the first two terms are not required in retrieving $\Delta\Phi(t)$, but $b_2$ and $b_3$ need to be characterized for SSSI operation.

In Eq.~\eqref{Eq2}, it is important to note that the modulation $\Delta\Phi(t)$ remains unchanged even if the time delay $\tau$ changes. This is because the term $e^{-i\omega\tau}$ shifts the modulation occurring at $t-\tau$ back to $t$. In other words, $\Delta\Phi(t)$ must be uniquely retrieved from many different $\tau$ delayed shots if the spectral phase $\varphi$ is correctly characterized.

\begin{figure}[htbp]
	\centering
	\includegraphics[width=0.9\linewidth]{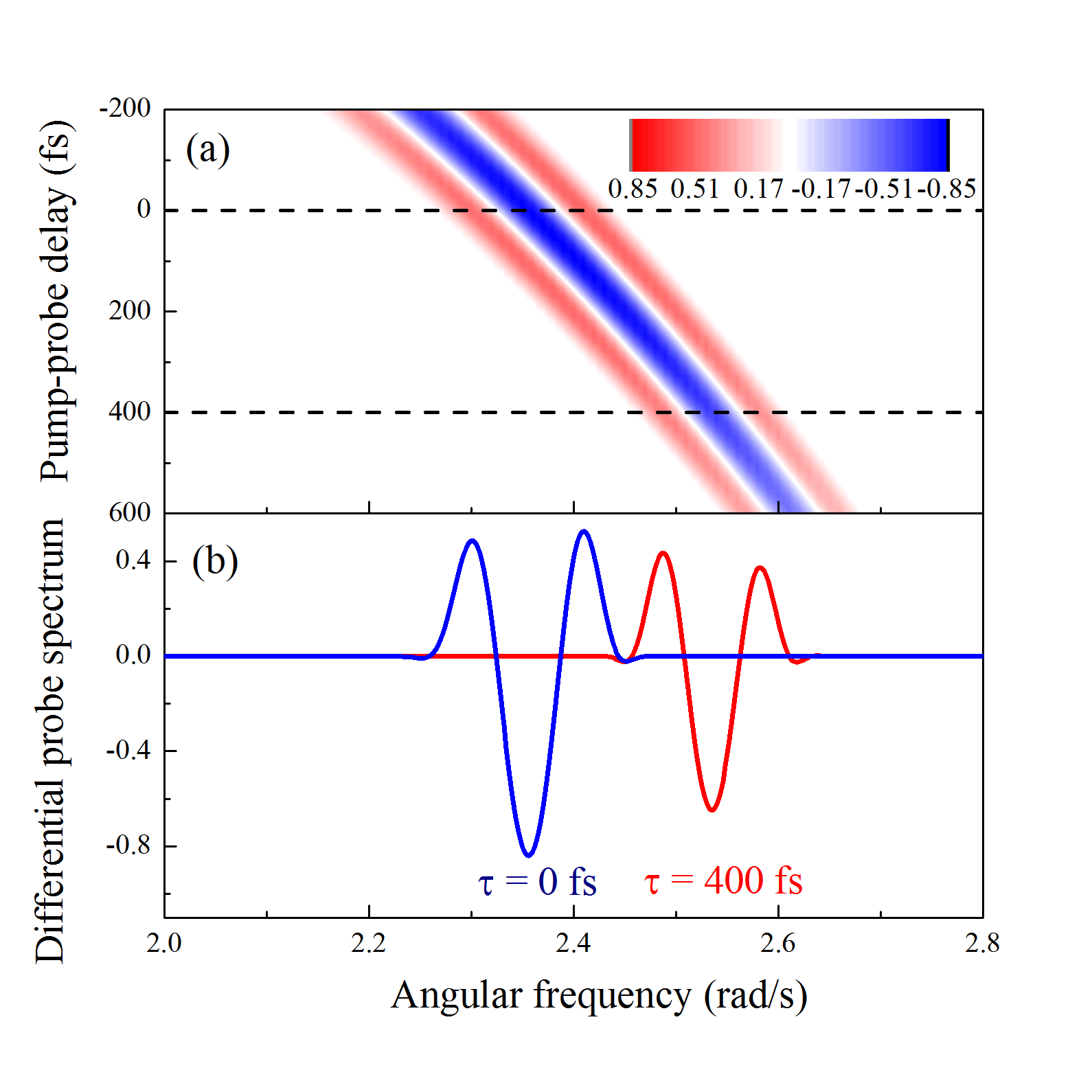}
		\caption{(a) Simulated differential probe power spectrum, $\Delta I(\omega,\tau)$, modulated by a phase transient given by Eq.~\eqref{Eq5} as a function of pump-probe delay. The probe is chirped with $b_2 = 1000~\text{fs}^2$ and $b_3 = 400~\text{fs}^3$.  (b) Differential probe spectral line-outs at $\tau = 0$~fs and $\tau=400$~fs }\label{Fig2}
\end{figure}

For illustration, we consider a Gaussian-type phase modulation given by 
\begin{equation}\label{Eq5}
 \Delta\Phi(t-\tau)=Ae^{-(t-\tau)^2/(2\sigma^2)},
\end{equation}
where we choose $A = 0.4$ and  $\sigma= 50~\text{fs}$. Here the probe pulse is also a Gaussian pulse centered at 800~nm with a full-width-at-half-maximum (FWHM) bandwidth of 170~nm and  chirped with $b_2 = 1000~\text{fs}^2$ and $b_3 = 400~\text{fs}^3$.  Figure~\ref{Fig2}(a) shows a series of differential probe power spectrum \cite{re17}, $\Delta I(\omega)$, as a function of the pump-probe delay $\tau$. Figure~\ref{Fig2}(b) shows two spectral line-outs at $\tau=0$~fs and 400~fs. One prominent feature is that the position of the central minimum $\omega_0$ shifts with respect to the pump-probe delay $\tau$. Here the central minimum is defined as the point where $\Delta I(\omega)$ oscillates slowest; mathematically, it is given by the condition
\begin{equation}
	\varphi'(\omega_0)=2b_2(\omega_0-\omega_c)+3b_3(\omega_0-\omega_c)^2=\tau.
\end{equation}
Therefore, by tracing $\omega_0$ at each $\tau$, one can determine $b_2$ and $b_3$ with a polynomial fit \cite{re17}. This method, however, is limited by the validity of the stationary phase approximation. Moreover, it is inefficient as only the central minimum/ maximum point or at most some adjacent extrema are used in each shot for characterization.

\begin{figure}[htpb]
	\centering
	\includegraphics[width=1\linewidth]{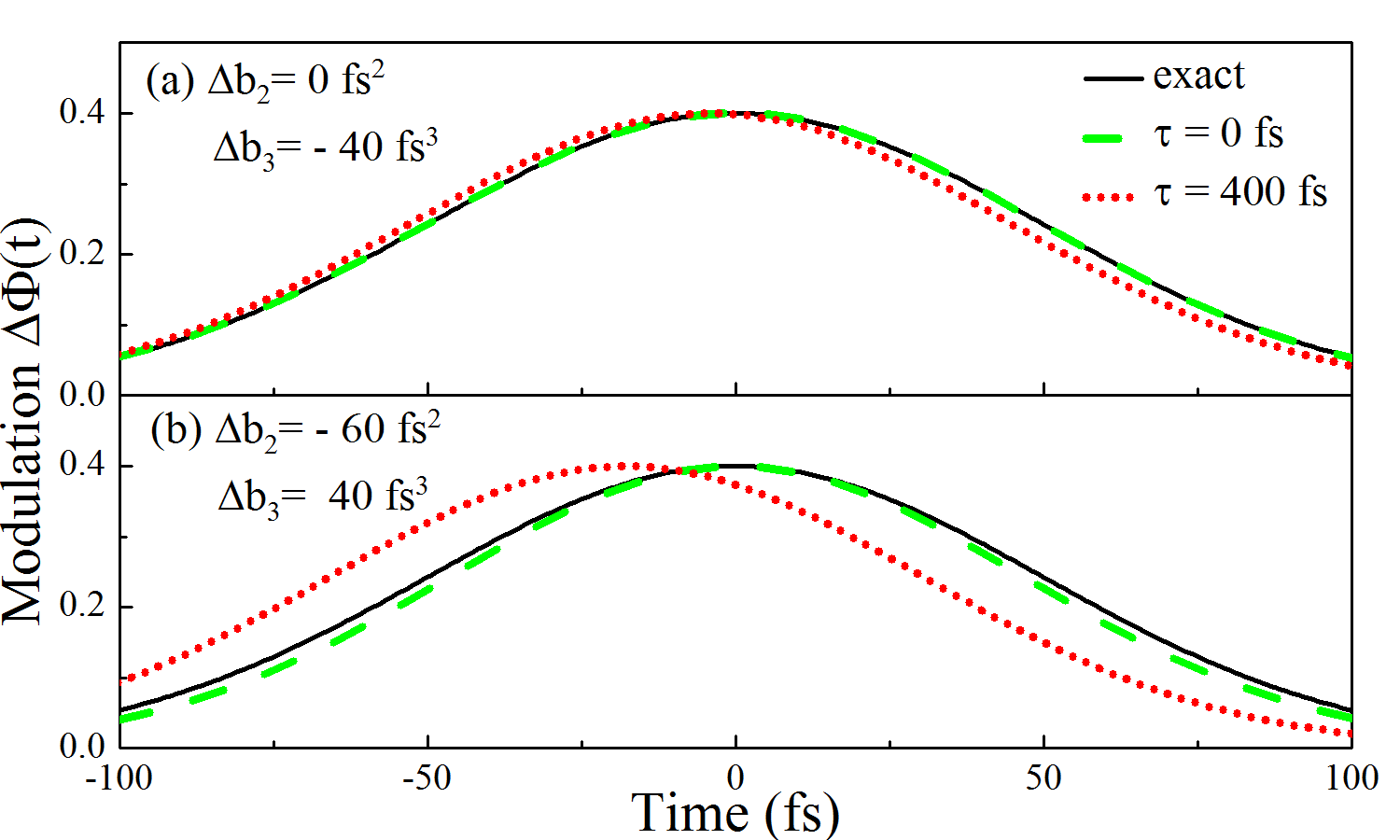}
	\includegraphics[width=1\linewidth]{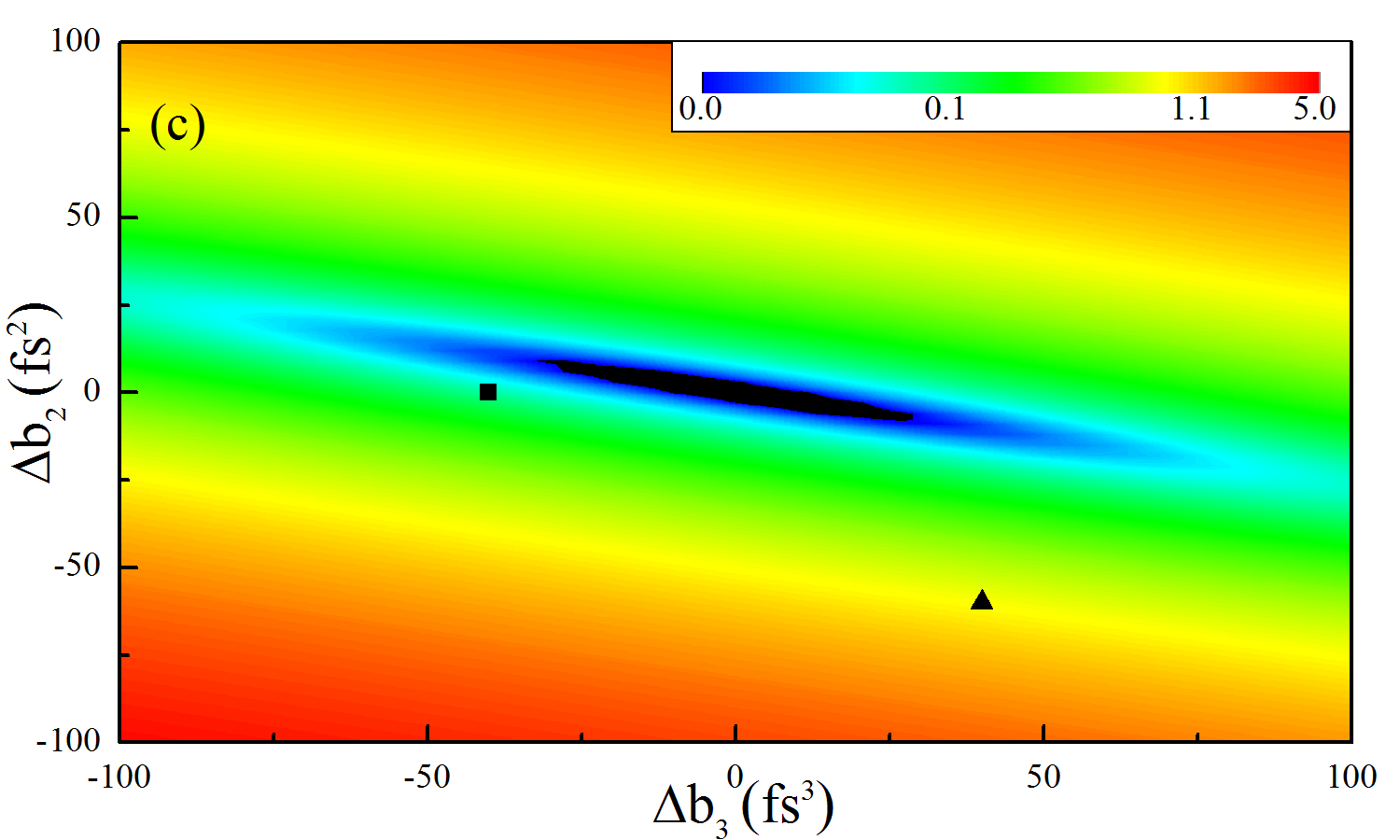}
	\caption{(a), (b) Extracted modulations $\Delta\Phi(t)$ obtained from two different time-delayed shots at $\tau=0$~fs (green dashed line) and $\tau=400$~fs (red dotted line) using intentionally incorrect spectral phase coefficients, along with the correct modulation (black line). The modulations $\Delta\Phi(t)$ obtained from two shots of different time delay are nonidentical when $\Delta b_2$ or $\Delta b_3$ is non-zero. (c) The dependence of $\Delta S^2$ on $\Delta b_2$ and $\Delta b_3$ shows a deep global minimum located at (0, 0). The square and the triangle correspond to $(\Delta b_2, \Delta b_3)=(0, -40)$ and $(-60, 40)$ as illustrated in (a), (b) respectively.}
	\label{Fig3}
\end{figure}

It is obvious that one needs the correct values of the second and third dispersion coefficients to properly characterize the phase modulation, but what is the consequence if the known values differ by $\Delta b_2$ and $\Delta b_3$ from the true ones? Our simulation shows that nonzero $\Delta b_2$ or $\Delta b_3$ lead to ambiguity in the retrieved modulation $\Delta\Phi(t)$. For illustration, we show the retrieved $\Delta\Phi(t)$'s from two different time delays $\tau=0$~fs and $\tau=400$~fs with $\Delta b_2=0$~fs$^2$, $\Delta b_3=-40$~fs$^3$ in Fig.~\ref{Fig3}(a) and $\Delta b_2=-60$~fs$^2$, $\Delta b_3=40$~fs$^3$ in Fig.~\ref{Fig3}(b). Those two retrieved $\Delta\Phi(t)$'s are different, and furthermore neither is identical to the true modulation. For a wider range of $\Delta b_2$ and $\Delta b_3$, we quantify the difference in shape of the retrieved modulations obtained from multiple time-delayed shots by
\begin{equation}\label{Eq6}
	\Delta {{S}^{2}}=\int_{-\infty }^{\infty }\sum\limits_{\tau }{{{{\left[ \Delta {{\Phi }_{\tau }}(t)-\overline{\Delta \Phi} (t) \right]}^{2}}dt}},
\end{equation}
where $\overline{\Delta \Phi} (t)$ is the average of $\Delta\Phi_\tau$ retrieved from all different time delays $\tau$. This $\Delta S^2$ strongly depends on how well the probe phase is characterized. For example, the dependence of $\Delta S^2$ on both $\Delta b_2$ and $\Delta b_3$ is computed and shown in Fig.~\ref{Fig3}(c). It clearly shows a deep global minimum located at $(0,0)$, which corresponds to the initially assigned probe phase ($b_2 = 1000$~fs$^2$ and $b_3 = 400$~fs$^3$). This shows that the modulations obtained from all different time delays converge only when the pulse's phase used for retrieval matches the true form.  At the same time, the converged function represents the real form of modulation. A mathematical proof of this observation is provided in Appendix.

\section{Algorithm details}
\noindent
Experimentally, it is possible that modulations from different shots are similar in shape but slightly different in magnitude due to pump pulse power fluctuation. Therefore, the modulation extracted from each shot is normalized prior to comparison. We also emphasize that the retrieved modulation often exhibits smooth variations in the vicinity of the central extremum, but it is very noisy in the far away region. Therefore, in practice, only a region of interest is used for an input. This should cover as much meaningful features of modulation as possible but be narrow enough to avoid too much noise.

One feature needs to be discussed is how to choose different time delays $\tau$  to optimize the operation of our algorithm. In a stationary phase approximation, the perturbed probe pulse can be expressed as \cite{re17}
\begin{equation}\label{Eq7}
\begin{split}
   \overline{E}(\omega )= &E(\omega )-C\frac{\Delta \Phi (\omega -{{\omega }_{0}})}{\sqrt{{{b}_{2}}'}}\left| E(\omega ) \right| \\
   & \quad \times \exp \left[ i{{b}_{2}}'{{(\omega -{{\omega }_{0}})}^{2}}+i{{b}_{3}}{{(\omega -{{\omega }_{0}})}^{3}} \right],
\end{split}
\end{equation}
where $C$ is a constant, $\Delta \Phi(\omega)$ is the Fourier transform of $\Delta \Phi(t)$ in the frequency domain, ${{b}_{2}}'={{b}_{2}}+3{{b}_{3}}({{\omega }_{0}}-{{\omega }_{c}})$, and $\omega_0$ is given by $\varphi '({{\omega }_{0}})=\tau $. In the case of a small $\tau$, $\omega_0$ becomes close to $\omega_c$ and ${{b}_{2}}'\approx {{b}_{2}}$, and the dominant part containing the third-order dispersion $3b_3(\omega_c-\omega_0)(\omega-\omega_0)^2$ becomes insignificant. In that case, the third order dispersion is hard to be determined. Therefore, for an effective operation, we want the change caused by the third-order dispersion to be greater than its measurement error $\varepsilon$
\begin{equation}
	\frac{3{{b}_{3}}({{\omega }_{0}}-{{\omega }_{c}})}{{{b}_{2}}}>\varepsilon ,
\end{equation}
where ${{\omega }_{0}}-{{\omega }_{c}}\approx \tau /(2{{b}_{2}})$. Therefore, the time delay separation between two shots should be
\begin{equation}\label{Eq8}
	\Delta \tau >\frac{\varepsilon b_{2}^{2}}{6{{b}_{3}}}.
\end{equation}
Equation~(\ref{Eq8}) establishes the relation between the time delay and experimental conditions. Furthermore, the upper limit of the time delay is fundamentally set by the probe pulse duration.

\begin{figure}[htbp]
	\centering
	\includegraphics[width=1\linewidth]{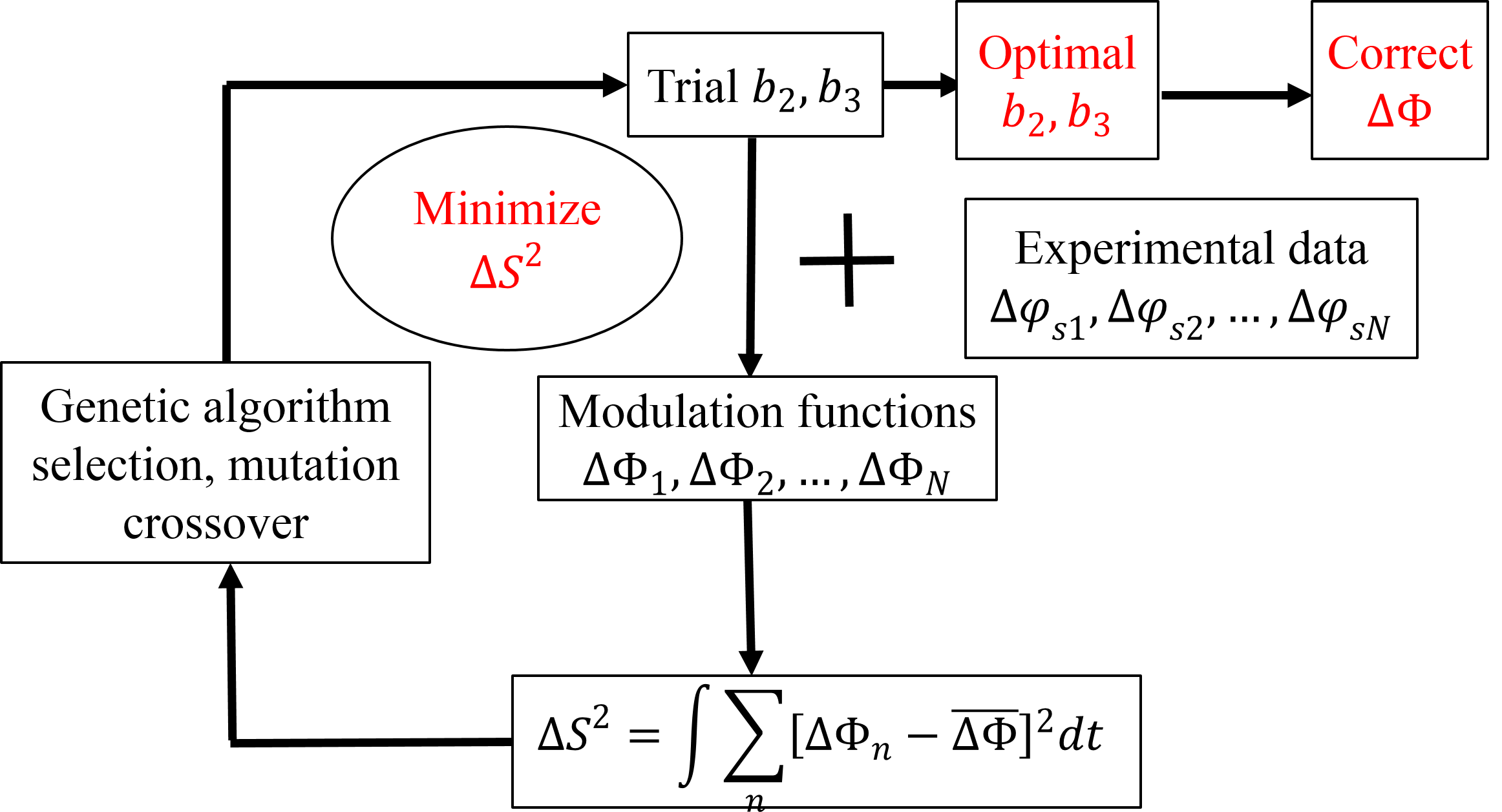}
	\caption{Algorithm routine for simultaneous characterization of both probe chirp ($b_2$ and $b_3$) and modulation ($\Delta\Phi(t)$). It uses a genetic algorithm (GA) to minimize $\Delta S^2$ such that the modulation functions obtained from all pump-probe delays can converge to equality.}
	\label{Fig4}
\end{figure}

As illustrated in Fig.~\ref{Fig3}, the probe spectral phase can be found by minimizing $\Delta S^2$. This process can be performed by a genetic algorithm (GA). Figure~\ref{Fig4} shows a diagram of our algorithm routine to characterize both probe and modulation simultaneously. First, the spectral modulations of probe $(\Delta \varphi_{s1},\Delta \varphi_{s2},..,\Delta \varphi_{sN})$ are measured at multiple pump-probe time delays $\tau_i$. Along with an initial population of $b_2$'s and $b_3$'s, the corresponding temporal modulation functions $(\Delta \Phi_{1},\Delta \Phi_{2},..,\Delta \Phi_{N})$ are constructed within Eq.~\eqref{Eq2}. Then the GA is used to minimize $\Delta S^2$ defined by Eq.~\eqref{Eq6}. Finally, those $b_2$, $b_3$ that provides the global minimum of $\Delta S^2$ will be selected for the best fitting parameters. Simultaneously, the $\overline{\Delta \Phi} (t)$ calculated from these optimized values is deemed to be the correct form of the applied modulation. 
\section{Performance test}
\noindent
In this section, we test the reliability of our algorithm with numerical simulations. Here we simulate two types of modulations. The first one mimics a Kerr-induced refractive index modulation, where the modulation is proportional to the intensity of a co-propagating pump pulse. The second one simulates a femtosecond photo-ionization process, where the modulation asymptotically approaches zero at $t \to -\infty$ and a non-zero value at $t \to \infty$. In both cases, the probe pulse is set to be the same as in the previous sections with $b_2 = 1000$~fs$^2$ and $b_3 = 400$~fs$^3$.
\begin{figure}
	\centering
	\includegraphics[width=1\linewidth]{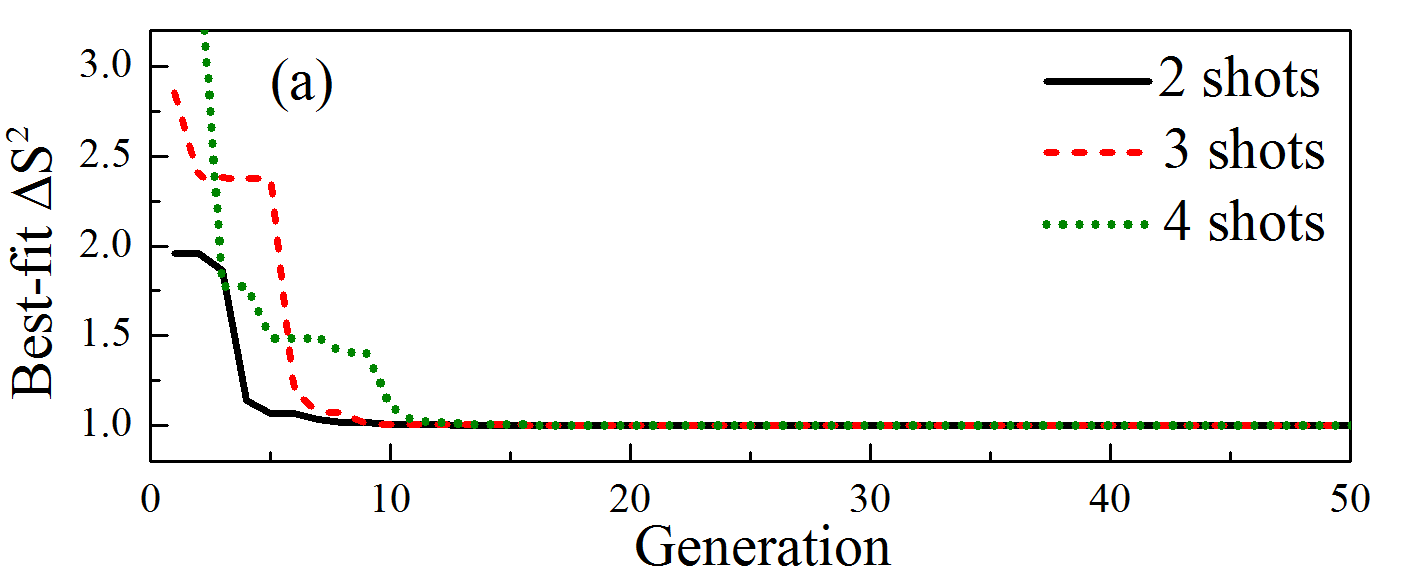}  	
	\includegraphics[scale=0.24]{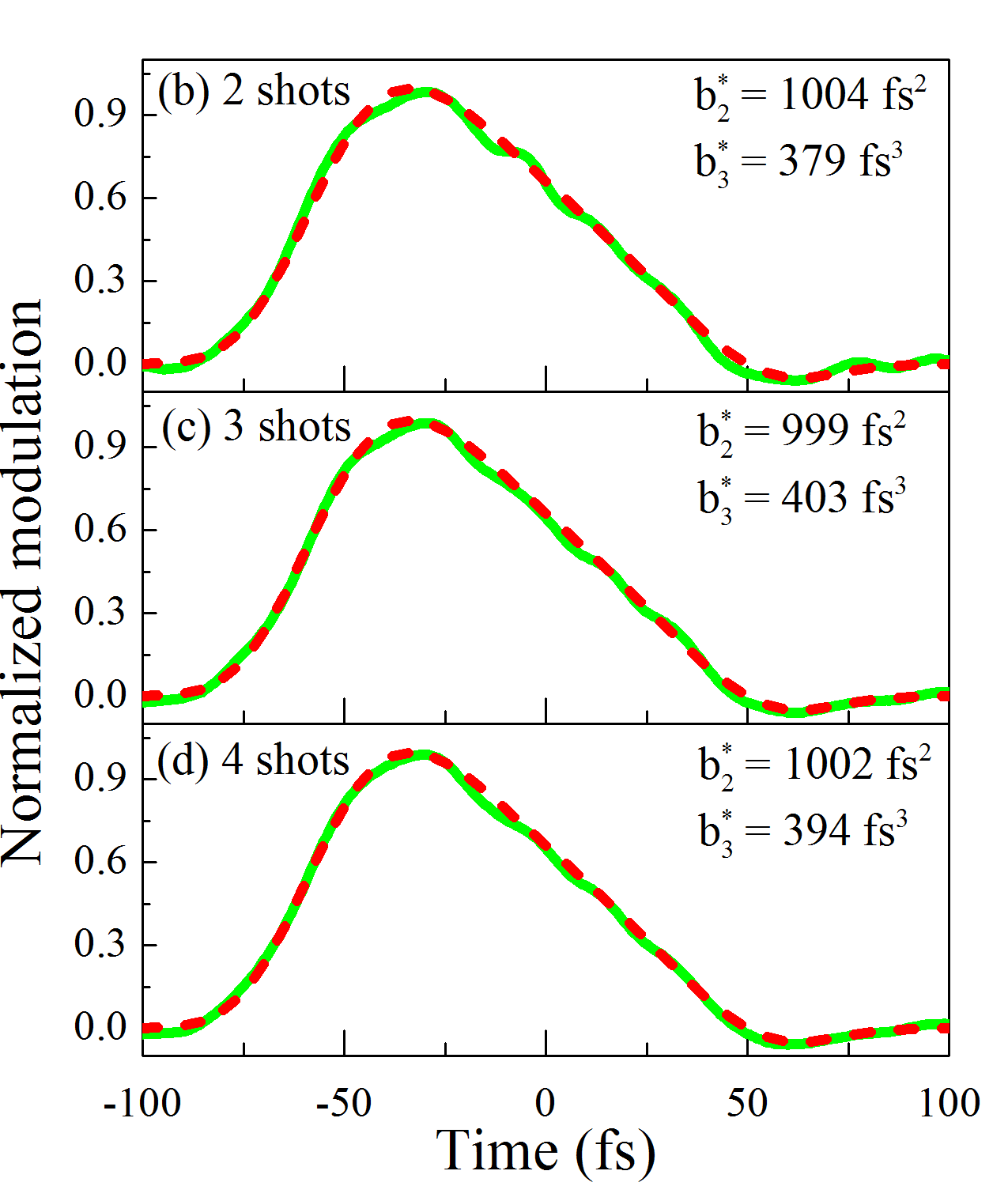}
	\includegraphics[scale=0.24]{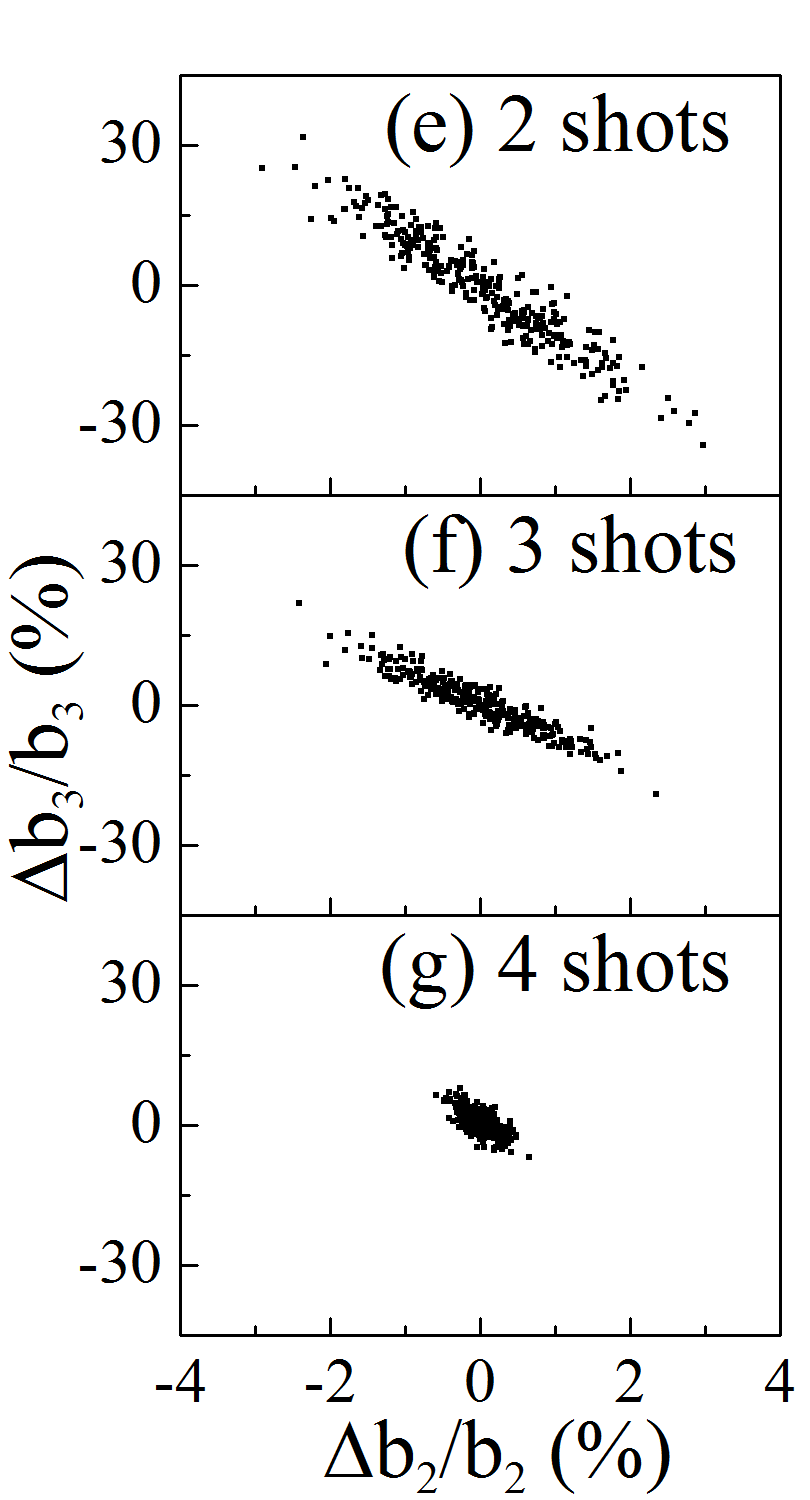}  
	\caption{(a) Minimal $\Delta S^2$ after each generation when using data from two shots at $\tau = 0$ and $400$~fs (black solid line), three shots at $\tau$ = 0, 400 and 600~fs (red dashed line), and four shots at $\tau$ = 0, 400, 600 and -200~fs (green dotted line). $\Delta S^2$ is normalized to the final converged value. (b)-(d) The real part of reconstructed average modulation (green solid line) compared to the exact function (red dashed line) given by Eq.~\eqref{Eq10}, with its best fitting parameters ($b_2^* \text{ and } b_3^*$) obtained from 2, 3, and 4 shots, respectively, as defined in (a). (e)-(g) Distribution of retrieved ($b_2,b_3$) after 300 trials, corresponding to (b-d) respectively.}
	\label{Fig5}
\end{figure}
\subsection{Kerr-like Modulation}\label{sec4a}
\noindent
An intense laser pulse can induce a transient in the index of refraction of a medium it propagates through, leading to a phase modulation on the co-propagating probe pulse. We assume the modulation has a form of
\begin{equation}\label{Eq10}
	\Delta \Phi (t)={{A}_{1}}{{e}^{-a{{t}^{4}}}}\left[ 1-b(t-{{t}_{0}}) \right]+i{{A}_{2}}{{e}^{-c{{t}^{2}}}}, 
\end{equation}
with $A_1=0.2, a=1.6\times 10^{-7}~\text{fs}^{-4}$, $b=5\times 10^{-2}~\text{fs}^{-1}$, $t_0=30~\text{fs}$, $A_2=0.1$, and $c=2.5\times 10^{-3}~\text{fs}^{-2}$. The imaginary part (second term) represents nonzero absorption in the medium. Here we assume the modulation is not symmetric in time. We also introduce a random error of $\le$5\% to the simulated spectrogram to test the stability of our algorithm and a fluctuation of $\le$10\% to the magnitude of $\Delta\Phi$ for each time-delayed shot.

We first test the convergence speed of our GA. In this simulation, each generation comprises 80 sets of $b_2, b_3$ with the search range of $600-1200$~fs$^2$ for $b_2$ and $200-800$~fs$^3$ for $b_3$. We perform the simulation in three cases with different numbers of time-delayed shots, namely two shots at $\tau = 0$ and $400$~fs, three shots at $\tau$ = 0, 400 and 600~fs, and four shots at $\tau$ = 0, 400, 600 and -200~fs. In each case, we use the same initial population that is intentionally chosen to be far away from the converged values. The optimized $\Delta S^2$ at each generation is shown in Fig.~\ref{Fig3}a. Despite the unfavorable condition we set, $\Delta S^2$ converges fast in all three cases after 15 generations.

In Figs.~\ref{Fig5}(b)-\ref{Fig5}(d), we show the optimized set of $b_2^*, b_3^*$, and the average modulation $\overline{\Delta\Phi}(t)$ obtained from two, three, and four delayed shots. In this example, the second order dispersion coefficient $b_2$ can be characterized within a 1\% error, and the shape of modulation can be reconstructed fairly well even with 2 shots. However, the third order dispersion coefficient $b_3$, less significant compared to $b_2$, suffers from a 5\% error when only two shots are used only. It is noticeable that when 3 and 4 shots are used, the retrieval errors of $b_2$ and $b_3$ reduce to less than 1\% and 2\%, respectively. 

We emphasize that the GA is so effective that it converges quickly to the almost exact global minimum of $\Delta S^2$ regardless of the number of shots used. Note that we also introduce $\le$10\% fluctuations to the modulation amplitude, but it is neutralized by the normalization step in our algorithm. Therefore, the retrieval error as shown previously is solely due to the random error introduced to the spectrogram. To examine how this error affects the retrieved values, we repeat the same simulation for 300 times and plot the extracted $b_2$ and $b_3$ in Figs.~\ref{Fig5}(e)-\ref{Fig5}(g). Firstly, compared to the typical uncertainty in $b_2$ (2\%), $b_3$ spans much wider with the standard deviation of $\sim$10\% in the case of 2 shots. This is understandable as the effect of third order dispersion on the spectra is quite small and can be overwhelmed by the random noise. Secondly, the overall certainty is diminished by increasing the number of snapshots. The error margins of retrieved $b_2$ and $b_3$ shrink significantly when the number of shots increase from 2 to 4, specially from $\sim$10\% to $\sim$2\% for $b_3$.  This is not surprising as the effect of random noise can be lessened by repetition. We note that our GA can always retrieve the exact $b_2$ and $b_3$, and $\Delta \Phi(t)$ when no random fluctuation is included in the simulations.
\subsection{Femtosecond Stepwise Modulation}
\begin{figure}
	\centering
	\includegraphics[width=0.9\linewidth]{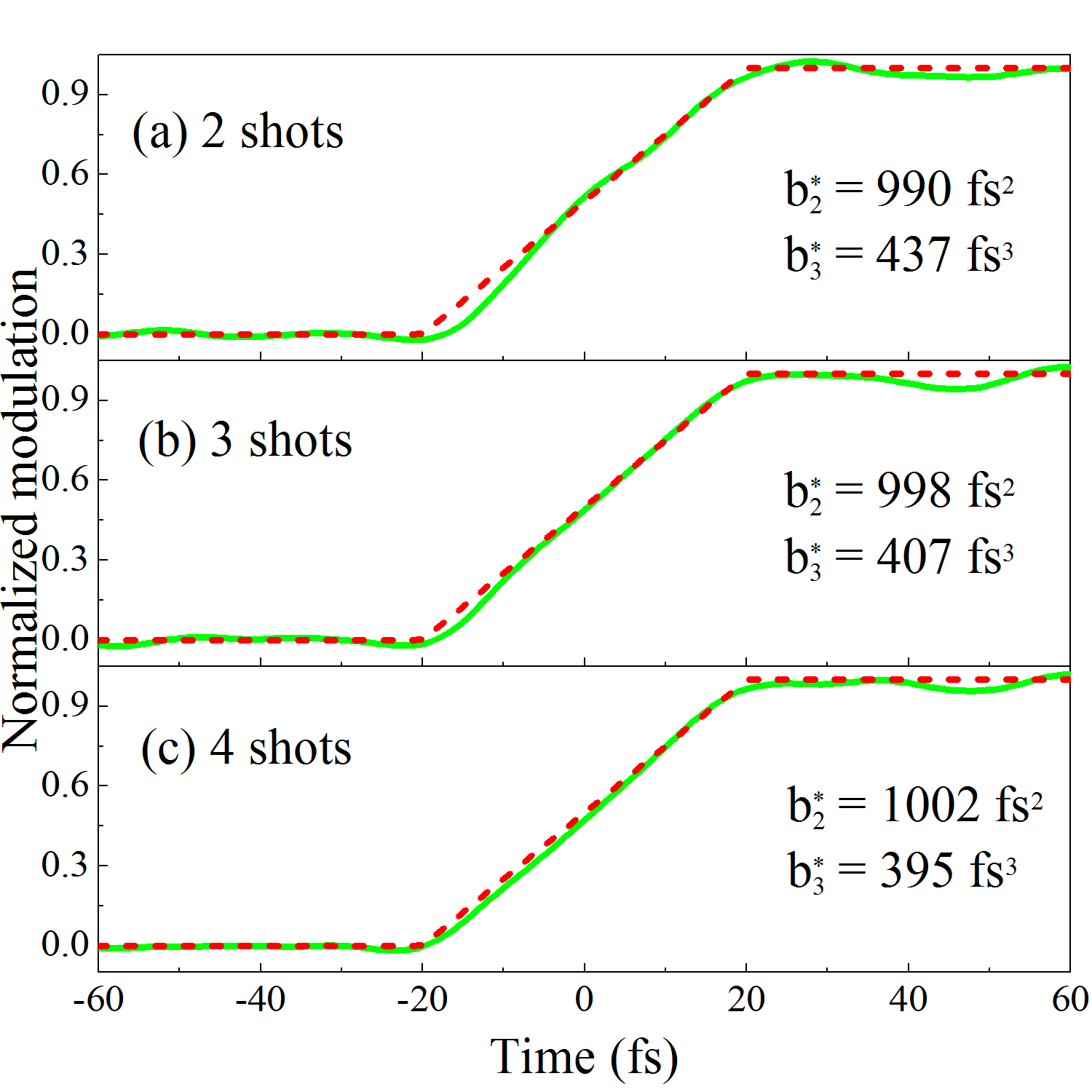}
	\caption{Same as Figs.~\ref{Fig5}(b)-\ref{Fig5}(d) but the modulation function is given by Eq.~\eqref{Eq11}.}
	\label{Fig6}
\end{figure}

\noindent
As a second example, we consider an ultrafast transient commonly observed in optical field ionization. In strong laser electric fields, atoms and molecules can be tunnel ionized, producing free electrons in continuum states. Macroscopically, the free electron density rises in time until the intense pulse passes by. The density modulation induced by the pump pulse can be picked up by a co-propagating probe pulse. For simplicity, we consider the following phase modulation caused by tunneling ionization,
\begin{equation}\label{Eq11}
	|\Delta \Phi (t)|=\left\{ \begin{matrix}
	0 & t\le-20~\text{fs} \\
	0.1(t+20) & -20~\text{fs}< t \le 20~\text{fs}  \\
	0.4 & t>20~\text{ fs}  \\
	\end{matrix} \right..
\end{equation}
Similar to the previous section, we simulate the spectrograms at different time delays and use data from 2, 3 or 4 shots to reconstruct the modulation. The spectrograms are also subject to $\le$5\% random fluctuations.  Our simulation results are presented in Figs.~\ref{Fig6}(a)-\ref{Fig6}(c). In the 2-shot case, the optimized $b_2^*$ and $b_3^*$ exhibit 1\% and 10\% errors, respectively. With the FWHM bandwidth of 170~nm, the fastest resolvable phase transient is $\sim$5.5~fs for a Gaussian temporal modulation. For an abruptly changing function as in this example, its Fourier transform spreads much wider in the frequency domain than that of a Gaussian function (for instance $1/|\omega|$  for a step function in time). This leads to an even worse temporal resolution. Thus a relatively large uncertainty is expected in the extraction process. However, when using three or four shots, highly accurate characterization is possible.

In conclusion, our algorithm works well for two examples of ultrafast modulations even with $\le$5\% random noise applied in the spectrograms. It will work equally well, we believe, for any reasonably shaped modulations. However, depending on the modulation shape, more than 2 shots are needed to obtain very high accuracy, especially when non-negligible random errors are present.

\subsection{Extending the Number of Fitting parameters}
\noindent
In this section, we test the flexibility of the algorithm when more fitting parameters are introduced. In one possible scenario, the probe and reference pulses can be non-identical with different $b_2$ and/or $b_3$. This disparity can occur when the reference and probe pulses pass through a beam splitter different numbers of times, thus leading to unequal dispersion. In that case, Eq.~\eqref{Eq2} needs to be modified as 
\begin{equation}
	\Delta \Phi (t)=-i\ln \left[ \frac{F\left\{ |\overline{E}(\omega )|e^{ i(\Delta {{\varphi }_{\tau }}+\varphi -\omega \tau) } \right\}}{F\left\{|E(\omega )|e^{ i(\varphi +\delta \varphi -\omega \tau)}\right\}} \right],
\end{equation}
where $\delta \varphi$ is the phase difference between the probe and reference pulses. We estimate 
\begin{equation}
	\delta\varphi \approx {{B}_{2}}{{(\omega -{{\omega }_{c}})}^{2}},
\end{equation}
where $B_2$ has an order of  10 fs$^2$. In this example, there are three parameters to be optimized ($b_2, b_3$ and $B_2$). Here we choose $b_2=1000$~fs$^2$, $b_3=400$~fs$^3$, and $B_2=30$~fs$^2$, with the same modulation and noise ($\le$5\%) as in Section \ref{sec4a} for simulation. The optimal parameters $b_2, b_3$ and $B_2$ retrieved from 2, 3 and 4 time-delayed shots are presented in Tab.~\ref{Tab1}. 

\begin{table}[htbp]
	\centering
	\caption{ Best-fit parameters ($b_2^*$, $b_3^*$, and $B_2^*$) retrieved with 2, 3, and 4 time-delayed shots when the probe and reference pulses are allowed to have second order dispersion coefficients different by $B_2$.}
	\begin{tabular}{cccc}
		\hline
		 & $b_2^*$ (fs$^2$) & $b_3^*$ (fs$^3$) & $B_2^*$ (fs$^2$)\\
		\hline
		2 shots & 1000 & 389 & 30.5 \\
		3 shots	& 997 &	398 &	29.8 \\
	    4 shots &	998&	398&	29.8\\
		\hline
	\end{tabular}
	\label{Tab1}
\end{table}
As shown in Tab.~\ref{Tab1}, $b_2$ and $B_2$ can be determined within 2\% regardless of the number of shots. Noticeably, when 3 or 4 shots are used, all three parameters can be obtained within 1\% error. Note that $b_2$ is retrieved with a 3\% difference, which is comparable to the 5\% error obtained when the reference and probe pulses are set to be identical ($B_2 = 0$) in Section \ref{sec4a}. Therefore, the addition of more chirp parameters does not significantly affect the performance of our algorithm.

\section{Conclusion}
\noindent
In summary, we have presented a simple method to determine both probe spectral phases and pump-induced modulations in a conventional SSSI setup. Our GA-based routine is shown to work for typical ultrafast modulations and capable of characterizing the probe phase with high accuracy. Also our algorithm can be easily modified to include more chirp parameters if necessary. With numerical simulations, we show that our algorithm is robust against random errors ($\le$5\%) and can provides satisfactory accuracy even with 2 time-delayed shots. With three or more shots, our algorithm can retrieve nearly the exact modulation and spectral phase. We believe our technique can be readily applied to any SSSI setup to simplify or eliminate its routine chirp characterization process.

\section*{Appendix: The uniqueness of $\mathbf{\Delta S^2}$  minimum}

\noindent
This section attempts to prove mathematically that the standard deviation $\Delta S^2$ exhibits a zero value only when the phase function used in extraction has the correct form. Suppose that $\varphi$ is slightly deviated as $\varphi \to \varphi + \delta\varphi$, then we have
\begin{align}
	 &\left| E(\omega ) \right|{{e}^{i(\varphi +\Delta \varphi )}}{{e}^{i\delta \varphi (\omega )}}{{e}^{-i\omega \tau }} \\
	 & \qquad ={{e}^{i\delta \varphi (\omega )}}{{e}^{-i\omega \tau }}F^{-1}\left[ E(t){{e}^{i\Delta \Phi (t-\tau )}} \right] \nonumber\\ 
	 &\qquad ={{e}^{i\delta \varphi (\omega )}}\int{M(\omega -{\omega }'){{e}^{i{\omega }'\tau }}\left| E(\omega ') \right|{{e}^{i\varphi ({\omega }')}}d{\omega }'}, \nonumber 
\end{align}\\
where $M(\omega)=F^{-1}[e^{i\Delta\Phi(t)}]$. The Fourier transform of this term is given by
\begin{align}\label{Eq16}
 & F\left\{ E(\omega ){{e}^{i(\varphi +\Delta \varphi )}}{{e}^{i\delta \varphi (\omega )}}{{e}^{-i\omega \tau }} \right\} \\ 
 & \propto \int{F\left[ \int{M(\omega -{\omega }')E({\omega }'){{e}^{i[\varphi ({\omega }')-{\omega }'\tau]}}d{\omega }'} \right]C(t'){{e}^{i\delta \tilde{\varphi }(t')}}dt'} \nonumber \\ 
 & \propto \int{{{e}^{i\Delta \Phi (t-{t}')}}|E(t-t'-\tau )|{{e}^{i\tilde{\varphi }(t-t'-\tau )}}C(t'){{e}^{i\delta \tilde{\varphi }(t')}}dt'},\nonumber
\end{align}
where $F\left\{ {{e}^{i\delta \varphi (\omega )}} \right\}=C(t){{e}^{i\delta \tilde{\varphi }(t)}}$ and $E(t)=|E(t)|{{e}^{i\tilde{\varphi }(t)}}$ . Generally, the modulation varies in a shorter time scale than the probe pule, so $\tilde{\varphi}$ and $\delta\tilde{\varphi}$ vary much faster than $\Delta\Phi$. Using the stationary phase approximation, Eq.~\eqref{Eq16} can be approximated as
\begin{align}
& F\left\{ E(\omega ){{e}^{i(\varphi +\Delta \varphi )}}{{e}^{i\delta \varphi (\omega )}}{{e}^{-i\omega \tau }} \right\} \\ 
& \propto {{e}^{i\Delta \Phi (t-g)}}\int{\left| E(t-t'-\tau ) \right|{{e}^{i\tilde{\varphi }(t-t'-\tau )}}C(t'){{e}^{i\delta \tilde{\varphi }(t')}}dt'},\nonumber
\end{align}
where $g$ is the point that contributes the most to the integral. Within the stationary phase approximation, this point can be given by
\begin{equation}\label{Eq18}
\delta \tilde{\varphi }'(g)-\tilde{\varphi }'(t-g-\tau )=0.
\end{equation}
As a result, Eq.~\eqref{Eq2} now gives
\begin{align}\label{Eq19}
	&\frac{F\left\{ E(\omega )\exp \left[ i(\Delta \varphi +\varphi +\delta \varphi ) \right]\exp (-i\omega \tau ) \right\}}{F\left\{ E(\omega )\exp (i\varphi +i\delta \varphi )\exp (-i\omega \tau ) \right\}} \\
	&\quad\approx \exp \left[ i\Delta \Phi (t-g) \right].\nonumber
\end{align}
Because $g$ depends on $\tau$ according to Eq.~\eqref{Eq18}, the extraction now produces different results at various time delays. Only when $\delta \varphi =0$, making $C(t)e^{i\delta\tilde{\phi}} \to \delta(t)$ and $g \to 0$, Eq.~\eqref{Eq19} yields the same function form regardless of $\tau$. 

\section*{Funding information}
\noindent
National Science Foundation (NSF) (Award No. 1351455); Air Force Office of Scientific Research (AFOSR) (FA9550-16-1-0163).

\bibliographystyle{apsrev}
\bibliography{reference}
\end{document}